\documentclass{article}

\usepackage[preprint]{neurips_2025}

\usepackage[utf8]{inputenc} 
\usepackage[T1]{fontenc}    
\usepackage{hyperref}       
\usepackage{url}            
\usepackage{booktabs}       
\usepackage{amsfonts}       
\usepackage{nicefrac}       
\usepackage{microtype}      
\usepackage{xcolor}         

\usepackage{graphicx}
\usepackage{amsmath}
\usepackage{amssymb}
\usepackage{bbding}
\usepackage{bigstrut}
\usepackage{multirow}
\usepackage{subcaption}
\usepackage{colortbl}
\usepackage{arydshln}
\usepackage{array}
\usepackage{wrapfig}

\title{Prediction Inconsistency Helps Achieve Generalizable Detection of Adversarial Examples}

\author{
Sicong Han\textsuperscript{1}, 
Chenhao Lin\textsuperscript{1}, 
Zhengyu Zhao\textsuperscript{1}, 
Xiyuan Wang\textsuperscript{2}, \\
\textbf{Xinlei He\textsuperscript{3}, 
Qian Li\textsuperscript{1}, 
Cong Wang\textsuperscript{4}, 
Qian Wang\textsuperscript{5}, 
Chao Shen\textsuperscript{1}} \\
\textsuperscript{1}School of Cyber Science and Engineering, Xi'an Jiaotong University, China \\
\textsuperscript{2}School of Automation Science and Technology, Xi'an Jiaotong University, China \\
\textsuperscript{3}DSA and IoT Thrust, HKUST (Guangzhou), China \\
\textsuperscript{4}Department of Computer Science, City University of Hong Kong, China \\
\textsuperscript{5}School of Cyber Science and Engineering, Wuhan University, China \\
\texttt{\{siconghan, zidonghuawxy\}@stu.xjtu.edu.cn}, \\
\texttt{\{linchenhao, zhengyu.zhao, qianlix, chaoshen\}@xjtu.edu.cn}, \\
\texttt{xinleihe@hkust-gz.edu.cn}, 
\texttt{congwang@cityu.edu.hk}, 
\texttt{qianwang@whu.edu.cn}
}

\begin{document}

\maketitle

\begin{abstract}
\label{sec:abstract}
Adversarial detection protects models from adversarial attacks by refusing suspicious test samples. However, current detection methods often suffer from weak generalization: their effectiveness tends to degrade significantly when applied to adversarially trained models rather than naturally trained ones, and they generally struggle to achieve consistent effectiveness across both white-box and black-box attack settings. In this work, we observe that an auxiliary model, differing from the primary model in training strategy or model architecture, tends to assign low confidence to the primary model's predictions on adversarial examples (AEs), while preserving high confidence on normal examples (NEs). Based on this discovery, we propose Prediction Inconsistency Detector (PID), a lightweight and generalizable detection framework to distinguish AEs from NEs by capturing the prediction inconsistency between the primal and auxiliary models. PID is compatible with both naturally and adversarially trained primal models and outperforms four detection methods across 3 white-box, 3 black-box, and 1 mixed adversarial attacks. Specifically, PID achieves average AUC scores of 99.29\% and 99.30\% on CIFAR-10 when the primal model is naturally and adversarially trained, respectively, and 98.31\% and 96.81\% on ImageNet under the same conditions, outperforming existing SOTAs by 4.70\%$\sim$25.46\%.
\end{abstract}

\section{Introduction}
\label{sec:introduction}
Deep neural networks (DNNs) have been proven to be vulnerable to adversarial examples (AEs)~\cite{goodfellow2014explaining,carlini2017towards}, which consist of carefully crafted imperceptible adversarial perturbations and normal examples (NEs). AEs are capable of misleading DNNs to output wrong predictions with high confidence regardless of original classification accuracy. Such a threat results in security concerns when deep learning technology is applied in safety-critical scenarios, e.g., medical image analysis systems~\cite{ma2021understanding}, autonomous driving systems~\cite{badjie2024adversarial}, etc. 

The countermeasures for protecting DNNs from being attacked by AEs can be roughly categorized into AE defense and AE detection. Defense techniques aim to enhance model robustness and enable correct prediction on AEs, with adversarial training~\cite{madry2017towards,wong2020fast} being a widely adopted approach.
However, its effectiveness remains limited, i.e., 65\% robust test accuracy is hard to achieve even for state-of-the-art adversarial trained models on ImageNet when confronted with strong adversarial attacks~\cite{liu2024comprehensive}.
Detection techniques attempt to predict whether the unknown input sample is AE or not and reject it if it is considered to be adversarial~\cite{xu2017feature,zhang2023detecting}. Such techniques are also an effective way to defeat adversarial attacks and are expected to work with defense techniques to promote model robustness and provide additional information on input samples.

Numerous AE detection methods have been proposed, and we divide them into two categories according to the required knowledge by the detector in this paper, i.e., \textit{white-box detection} and \textit{black-box detection}. White-box detection needs the full knowledge of the protected model, either to generate white-box AEs to train the detector or using the features extracted from the intermediate layer to distinguish between AEs and NEs~\cite{zhang2023detecting,tian2021detecting,ma2018characterizing,wang2023detecting}. Black-box detection treats the protected model as a black-box model and can only access to its output, which typically applies the input transformation or denoising techniques to disrupt adversarial perturbation, inducing prediction fluctuations that signal potential AEs~\cite{xu2017feature,tian2018detecting}.

Existing AE detection methods are predominantly white-box approaches, which achieve promising results but often suffer from high computational costs. In contrast, black-box detection, while easier to deploy, is less effective against stronger attacks. Moreover, both types of methods face a critical limitation in terms of generalization: (1) most existing detection methods are developed and evaluated on naturally trained models, and often exhibit poor compatibility with adversarially trained models; and (2) prior evaluations have primarily focused on white-box adversarial attacks, while it has been shown that some of detection methods perform poorly against black-box adversarial attacks~\cite{aldahdooh2022adversarial}. 

\begin{wrapfigure}{r}{0.5\textwidth}
  \centering
  \includegraphics[width=0.45\textwidth]{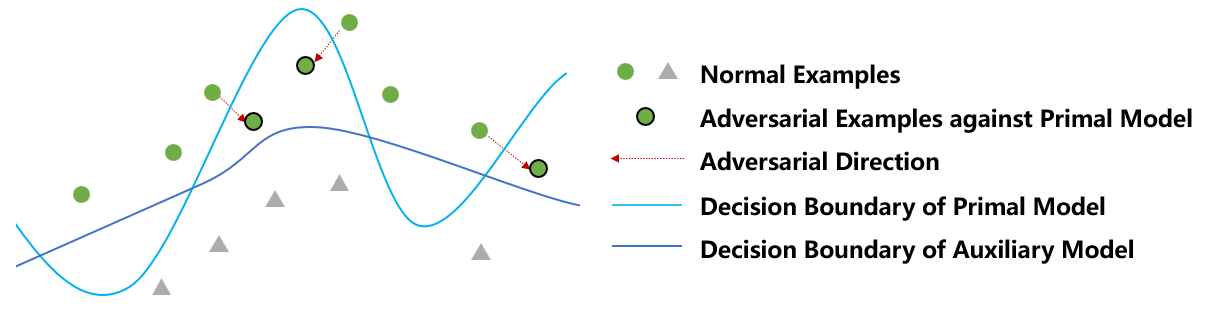}
  \caption{Adversarial examples against one (primal) model may not be adversarial against another (auxiliary) model. Such prediction inconsistency can be used to detect adversarial examples.}
  \label{fig:illustration}
\end{wrapfigure}
To mitigate the existing limitations, this paper proposes a black-box detection method exploiting the prediction inconsistency of models on AEs. 
We observe a clear discrepancy in prediction results from the primary and auxiliary models on AEs.
When the auxiliary model differs from the primal model, either due to variations in training methods or differences in model architectures, it becomes more challenging for AEs to cross the decision boundary of the auxiliary model, as illustrated in Figure~\ref{fig:illustration}. As a result, the auxiliary model tends to assign low confidence scores to the labels predicted by the primal model. In contrast, both primal and auxiliary models typically maintain consistent predictions on NEs.

This phenomenon, observed in both naturally and adversarially trained models, inspires us to design a simple yet flexible detection method named \textit{Prediction Inconsistency Detector} (PID). 
PID introduces an auxiliary model that differs from the primal model either in training approach or architecture, and four metrics are designed to measure the prediction inconsistency $I_{pred}$ between the two models to distinguish between NEs and AEs.
Among these metrics, the most effective one works as follows: first, the primal model assigns a predicted label to the test sample; then, the confidence score on the same label from the auxiliary model is acquired, which is then employed to calculate prediction inconsistency $I_{pred}$. The higher the $I_{pred}$, the more likely the test sample is an AE. 
The proposed PID can be complementary to both naturally and adversarially trained models to efficiently protect models from various adversarial attacks.
To summarize, our work makes the following contributions: 
\begin{itemize}
\item[$\bullet$]
We identify the significant prediction inconsistency between primal and auxiliary models on AEs, which forms the basis of our novel black-box detection method named PID. It exploits the hard label of the primal model and the soft label from the auxiliary model to calculate the prediction inconsistency and effectively detect AEs without requiring prior knowledge of the protected primal model.
\item[$\bullet$]
We design PID to work effectively with both naturally and adversarially trained models, making it more generalizable than existing methods that suffer from reduced effectiveness when applied to adversarially trained models.
\item[$\bullet$]
We conduct extensive experiments on 2 datasets and evaluate the detection performance of PID against 3 white-box, 3 black-box, and 1 mixed attacks with multiple perturbation sizes, where black-box attacks are often overlooked in previous works. Experimental results demonstrate PID's superior generalization ability in detecting different types and strengths of adversarial attacks.
\end{itemize}

\section{Related Work}
\textbf{Adversarial Attacks.} The aim of adversarial attacks is to add imperceptible perturbations to NEs to mislead DNNs. Based on the attackers' knowledge, adversarial attacks can be roughly divided into two categories, i.e., white-box adversarial attacks and black-box adversarial attacks.

White-box adversarial attacks require full knowledge of target models to craft AEs, which can expose the weakness of DNNs. 
On the basis of the Fast Gradient Sign Method (FGSM)~\cite{goodfellow2014explaining} attack, the Projected Gradient Descent (PGD) attack~\cite{madry2017towards} promoted attack performance by introducing random initialization and performing multi-step attacks along the perturbation direction. To eliminate the effects brought by the weaknesses in the standard PGD attack, e.g., fixed step size, Croce et al.~\cite{croce2020reliable} proposed Auto-PGD with two different loss functions and further combined it with FAB~\cite{croce2020minimally} and black-box Square~\cite{andriushchenko2020square} attacks to form a parameter-free ensemble attack named AutoAttack (AA). It is considered one of the strongest adversarial attacks.  
In addition, Carlini and Wagner (C\&W) attack~\cite{carlini2017towards} proposed a new object function, which was optimized by adopting general optimizers, e.g., Adam~\cite{kingma2014adam}. Moosavi et al.~\cite{moosavi2016deepfool} developed the DeepFool attack, which aimed to find small perturbations by leading the image toward the nearest decision boundary in each update. 

The assumption that attackers can acquire all the information of target models makes white-box adversarial attacks less practical. 
Numerous black-box adversarial attacks that require less knowledge have been proposed in recent years. 
Black-box adversarial attacks fall into three categories according to different levels of the required knowledge of target models.
Decision-based black-box attacks only rely on the predicted labels. Triangle Attack (TA)~\cite{wang2022triangle} optimized the perturbations exploiting the geometric information of the triangle constructed by a benign sample, the current and the next adversarial examples, operating in the low-frequency space to reduce queries.
Score-based black-box attacks utilize the scores predicted by the model to generate AEs. Square attack~\cite{andriushchenko2020square} achieves high query efficiency by combining a specialized initialization strategy with square-shaped update steps.
Transfer-based black-box attacks exploit the transferability of AEs. In~\cite{wang2021enhancing}, Wang and He enhanced the transferability of AEs by using gradient variance of the previous iterations, which was further combined with NI-FGSM attack~\cite{lin2019nesterov} to form the VNI-FGSM attack. 


\noindent \textbf{Detection of AEs.} In contrast to defense methods, which focus on enhancing robust test accuracy, detection methods identify and reject AEs by distinguishing them from NEs, thereby safeguarding the model.
Most detection methods treat the protected model as a white-box model, generating AEs against the model to train the detector or extracting the features from the intermediate layers to analyze the differences between AEs and NEs. 
Ma et al.~\cite{ma2018characterizing} extracted Local Intrinsic Dimensionality (LID) features of NEs and corresponding generated adversarial and noisy versions from each transformation layer to train a classifier to detect AEs.
Tian et al.~\cite{tian2021detecting} proposed the Sensitivity Inconsistency Detector (SID), which trained a dual classifier combined with the primal classifier to expose the sensitivity of AEs to the decision boundary fluctuation.
In~\cite{zhang2023detecting}, EPS-AD was developed, where the pre-trained diffusion model~\cite{song2021score} was adopted to estimate the expected perturbation score (EPS) of test samples, and EPS-based maximum mean discrepancy (MMD) was used as the metric to measure the discrepancy between NEs and AEs.
There are also some detection methods using the input transformation to disrupt the perturbations to cause the prediction fluctuation, which treats the protected model as a black-box model.
Xu et al.~\cite{xu2017feature} proposed Feature Squeezing (FS), a detection method that applies several feature squeezers to reduce the perturbation space. Inputs showing significant prediction inconsistency before and after squeezing are flagged as adversarial.

\textbf{This Work.} Most detection methods are designed for naturally trained models and often lose effectiveness when applied to adversarially trained ones. Given that adversarial training still struggles against large perturbations and black-box attacks, it is necessary to develop a detection method that can be integrated with both naturally and adversarially trained models. Additionally, current evaluations on detection methods remain incomplete due to the limited inclusion of diverse black-box attacks, leading to an inadequate assessment of the generalization and reliability of detectors.
To address these limitations, this work (1) proposes a detection method named PID that is compatible with both naturally and adversarially trained models, and (2) conducts evaluations using a comprehensive set of attacks, including three white-box attacks (PGD, C\&W, DeepFool), three black-box attacks (TA, Square, VNI-FGSM), and one mixed attack (AA), forming a more challenging test environment.

\section{Method}
\subsection{Motivation}
\label{sec:motivation}

\begin{figure}[t]
    \centering
    \begin{subfigure}{0.23\textwidth}
        \centering
        \includegraphics[width=\textwidth]{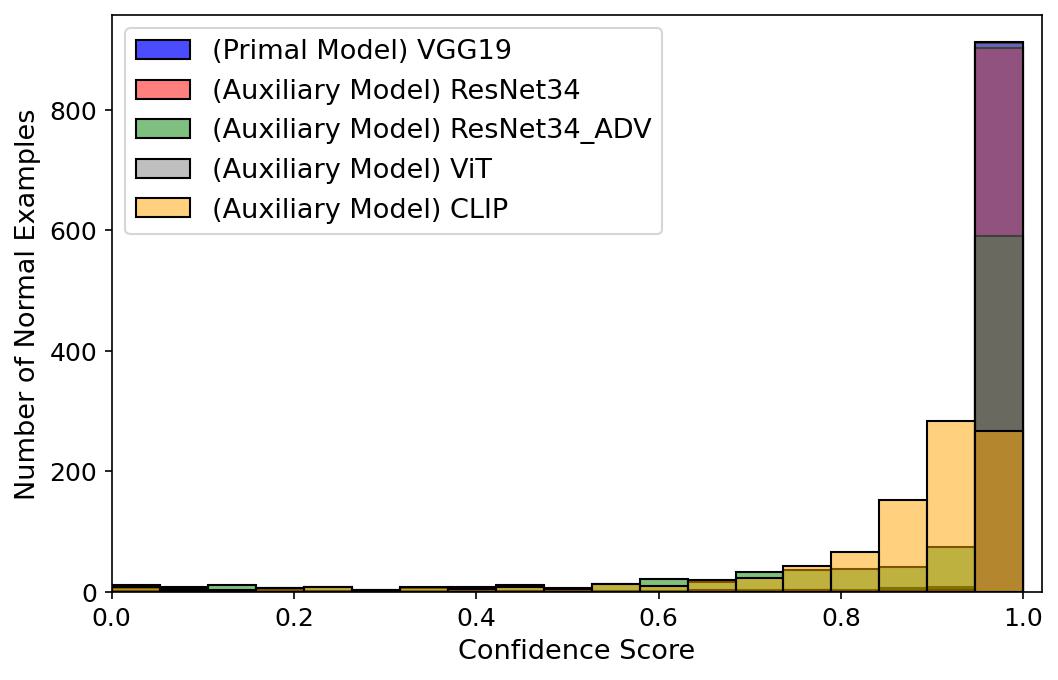}
        \caption{}
        \label{fig:vgg_clean}
    \end{subfigure}
    \begin{subfigure}{0.23\textwidth}
        \centering
        \includegraphics[width=\textwidth]{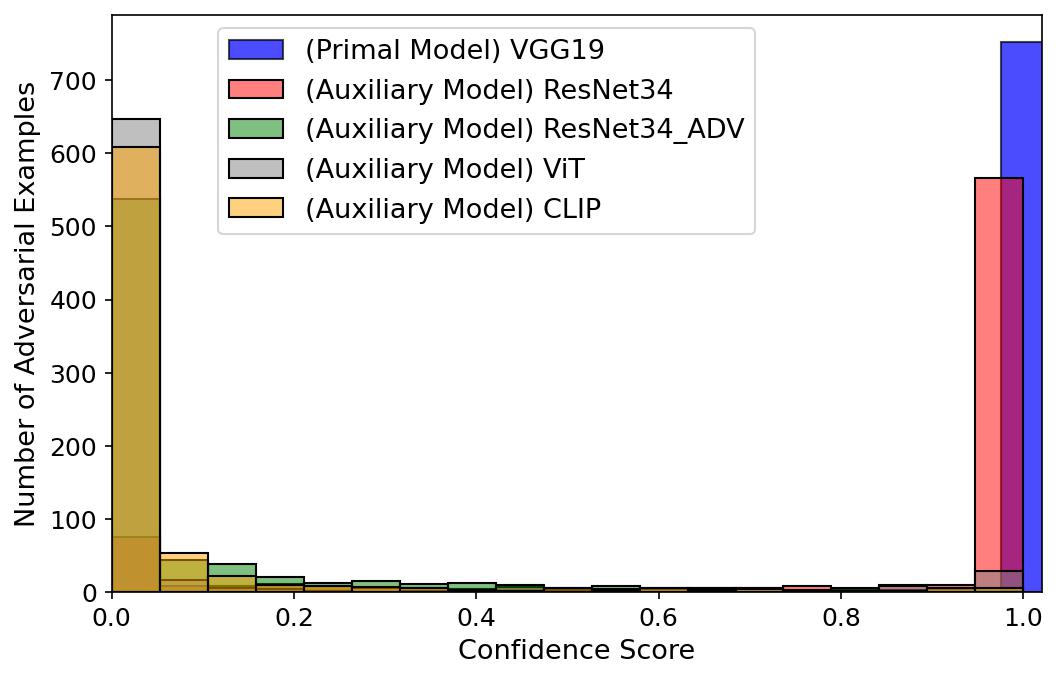}
        \caption{}
        \label{fig:vgg_adv}
    \end{subfigure}
    \begin{subfigure}{0.23\textwidth}
        \centering
        \includegraphics[width=\textwidth]{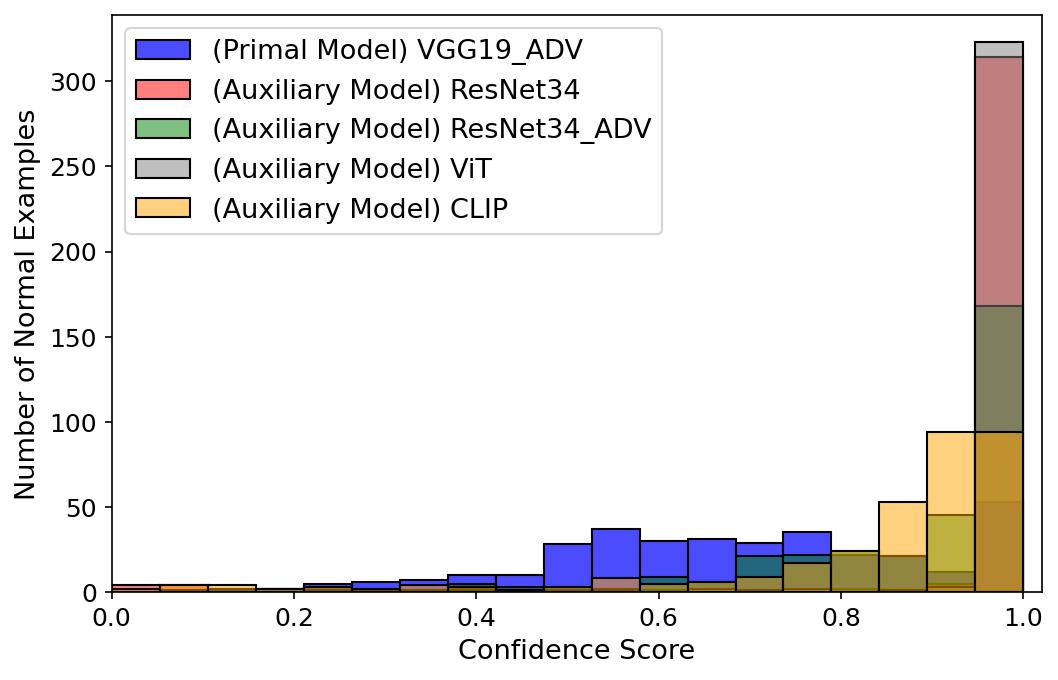}
        \caption{}
        \label{fig:vgg_at_clean}
    \end{subfigure}
    \begin{subfigure}{0.23\textwidth}
        \centering
        \includegraphics[width=\textwidth]{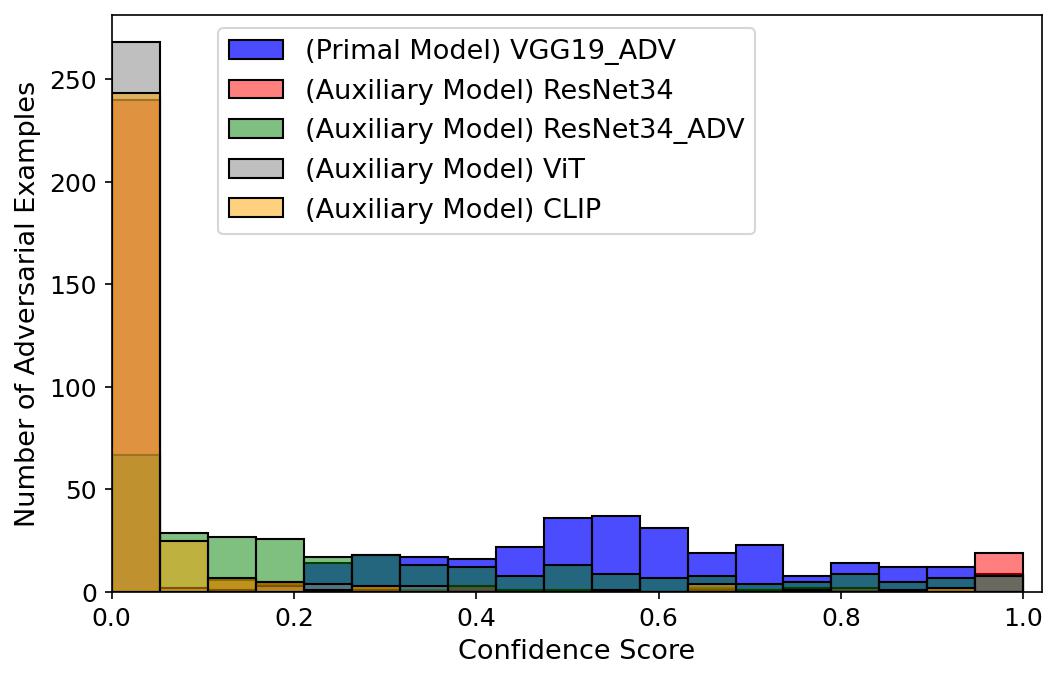}
        \caption{}
        \label{fig:vgg_at_adv}
    \end{subfigure}
    \caption{Confidence score distribution for \textit{“Airplane”} label across primal and auxiliary models on (a) NEs classified by the naturally trained VGG19, (b) AEs classified by the naturally trained VGG19, (c) NEs classified by the adversarially trained VGG19, (d) AEs classified by the adversarially trained VGG19, respectively.}
    \label{fig:motivation}
\end{figure}

It is pointed out in~\cite{tian2021detecting} that, AEs are sensitive to the fluctuation of the highly-curved region of the decision boundary, which can be exposed by training a dual classifier having dissimilar structures at the highly-curved regions with the original classifier while maintaining similar structures at the other regions.
Inspired by this phenomenon, we explore whether AEs are sensitive to the fluctuation of decision boundaries caused by introducing auxiliary models, either with different training methods or in different model architectures compared to the primal model.

Specifically, we adopt the naturally trained VGG19 (denoted as VGG19) as the primal model on CIFAR-10 and select NEs that are correctly classified as \textit{“Airplane”} by VGG19. Then four models are used as auxiliary models: the naturally trained ResNet34 (denoted as ResNet34), the adversarially trained ResNet34 (denoted as ResNet34\_ADV), the naturally trained Vision Transformer (ViT)-L/16~\cite{dosovitskiy2021image} (denoted as ViT), and the naturally trained Contrastive Language-Image Pre-Training (CLIP) model~\cite{radford2021learning} (denoted as CLIP). Each auxiliary model is used to classify these NEs, and the confidence scores for the \textit{“Airplane”} label serve as an intuitive indicator of sensitivity to decision boundary fluctuations, where lower scores indicate greater sensitivity.
Furthermore, we choose AEs (generated by PGD attack with perturbation size of 8/255) that are wrongly classified as \textit{“Airplane”} by VGG19 and use the same four auxiliary models to classify them and obtain the corresponding confidence scores for the \textit{“Airplane”} label.
The confidence score distributions for \textit{“Airplane”} label across primal and auxiliary models on NEs and AEs are depicted in Figure \ref{fig:vgg_clean} and \ref{fig:vgg_adv}, respectively. From these figures, we observe that for these NEs classified as \textit{“Airplane”} by the primal model, all four auxiliary models output high confidence scores on the same label, with the CLIP model being slightly less confident.
In contrast, for AEs labeled as \textit{“Airplane”} by VGG19, ResNet34\_ADV, ViT, and CLIP assign notably low confidence scores, while ResNet34 still gives high scores due to the transferability of AEs across Convolutional Neural Networks (CNNs).

Similar experiments are implemented using the adversarially trained VGG19 (denoted as VGG19\_ADV) as the primal model on CIFAR-10. NEs and AEs predicted as \textit{“Airplane”} by VGG19\_ADV are selected, and their confidence score distributions across primal and auxiliary models are presented in Figure~\ref{fig:vgg_at_clean} and Figure~\ref{fig:vgg_at_adv}, respectively. It can be observed that four auxiliary models still assign low confidence scores for the AEs labeled as \textit{“Airplane”} by VGG19\_ADV.
Furthermore, among the auxiliary models working with both naturally and adversarially trained models, ViT tend to assign the lowest confidence scores to the labels predicted by the primal model for AEs.

It is evident that AEs are sensitive to the decision boundary fluctuation caused by introducing auxiliary models—whether trained using different approaches (e.g., natural training vs. adversarial training) or featuring distinct architectures (e.g., CNN vs. ViT). 
This sensitivity manifests as a noticeable drop in the confidence scores assigned by the auxiliary model to the labels predicted by the primal model. As illustrated in Figure~\ref{fig:motivation}, this prediction inconsistency is minimal for normal examples (NEs) but pronounced for AEs, making it an intuitive and effective signal for detection.
This motivates us to design a detection method, which can work with both naturally and adversarially trained models, through introducing an auxiliary model and exploiting the model prediction inconsistency to tell NEs and AEs apart. The details of \textit{Prediction Inconsistency Detector} (PID) are given as follows.

\subsection{Design of PID}
\label{sec:design}

\noindent \textbf{Preliminaries.} Let $f(\cdot)$ denote the primal model, which is the $k$-class classifier, and its output can be hard label $y={\arg\max}_i\{f_i(x)\}$, where $f_i(x)\in[0,1]$ is the $i$-th class confidence score of the input $x$, and $i=1,2,\cdots,k$.
Since our detection method is designed to work with both naturally and adversarially trained models, the primal model $f(\cdot)$ can also be obtained through adversarial training and its variants~\cite{wong2020fast,jia2024improving}.
$g(\cdot)$ represents the other $k$-class classifier used as the auxiliary model, and it outputs the confidence score $g(x)=\{g_1(x),g_2(x),\cdots,g_k(x)\}$, where $g_j(x)\in [0,1]$ is the $j$-th class confidence score of the input $x$, and $j=1,2,\cdots,k$.

\begin{wrapfigure}{r}{0.5\textwidth}
  \centering
  \includegraphics[width=0.48\textwidth]{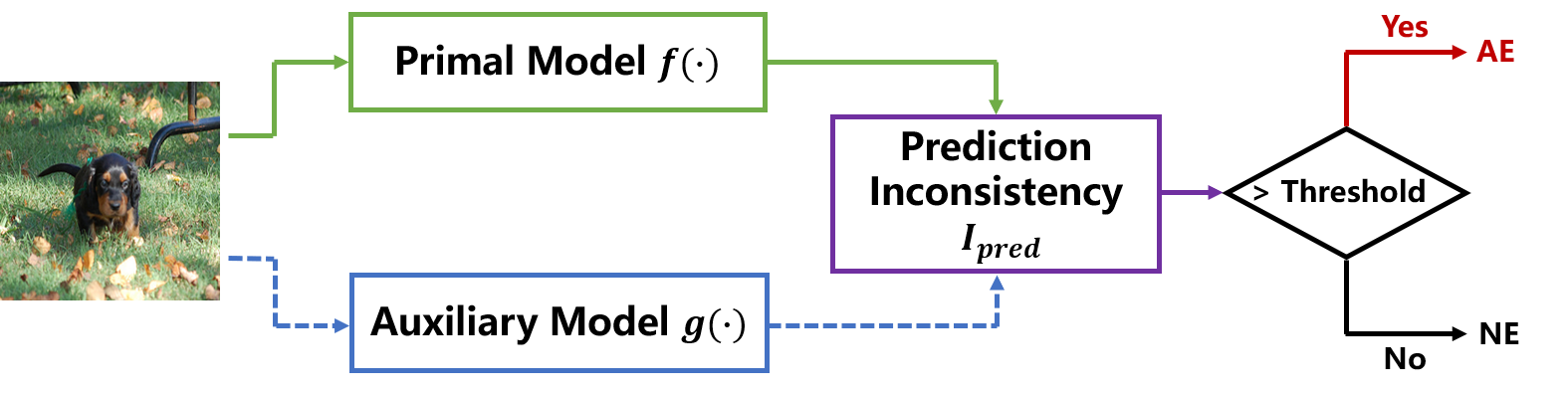}
  \caption{Overview of the proposed Prediction Inconsistency Detector (PID) for detecting AEs. The test sample are fed into primal and auxiliary models, respectively, to obtain the prediction inconsistency $I_{pred}$ by using the designed metrics. If $I_{pred}$ exceeds a threshold value, the test sample is judged to be adversarial.}
  \label{fig:detector}
\end{wrapfigure}
\noindent \textbf{Metrics for Quantifying Prediction Inconsistency.} The process of PID is illustrated in Figure \ref{fig:detector}, where 4 metrics are designed for measuring the prediction inconsistency.

\textit{Metric 1.} First, the test sample $x$ is input into the primal model $f(\cdot)$ and the assigned label $y$ can be acquired. Then $x$ is input into the auxiliary model $g(\cdot)$, and the confidence score $g_y(x)$ corresponding to label $y$ is obtained. Metric 1 for quantifying the prediction inconsistency $I_{pred}$ can be calculated as,
 \begin{gather}
     y={\arg\max}_i\{f_i(x)\}, \notag \\
     I_{pred}= 1 - g_y(x).
     \label{eq:prob}
 \end{gather}
If $x$ is an NE, $I_{pred}$ tends to be small (approaching 0), as a model trained on the same dataset is likely to classify it correctly with high confidence. Conversely, if $x$ is an AE, $I_{pred}$ can be large (approaching 1), especially when the attack fails to fool the auxiliary model $g(\cdot)$. Even when $x$ successfully misleads $g(\cdot)$, a low confidence score can still result in a high $I_{pred}$, making the AE more detectable.

Many detection methods~\cite{xu2017feature,tian2021detecting} leverage differences between soft labels to estimate the likelihood of a test sample being adversarial. Following this idea, we propose metrics based on the soft label discrepancies between the primal and auxiliary models to quantify prediction inconsistency and detect AEs. For the rest of other metrics described in this subsection, both $k$-class classifiers $f(\cdot)$ and $g(\cdot)$ output confidence scores, where $f(x)=\{f_1(x), f_2(x), \cdots, f_k(x)\}$, $g(x)=\{g_1(x), g_2(x), \cdots, g_k(x)\}$. 

\textit{Metric 2.} We define Metric 2 as follows,
 \begin{gather}
     y={\arg\max}_i\{f_i(x)\}, \notag \\
     I_{pred}= f_y(x) - g_y(x),
 \end{gather}
where $f_y(x)$ and $g_y(x)$ denote the confidence scores for label $y$ from $f(\cdot)$ and $g(\cdot)$, respectively.

\textit{Metric 3.} Following~\cite{xu2017feature}, prediction inconsistency can also be calculated using the $\ell_1$ norm of the difference between two prediction vectors. 
In Metric 3, we first sort confidence scores $f(x)$ in descending order and select the $n$ highest confidence scores $f(x)_{\text{top}-n}=\{f_{y_1}(x), f_{y_2}(x), \cdots, f_{y_n}(x)\}$ and corresponding labels $\{y_1, y_2,\cdots, y_n\}$. Then we obtain the confidence scores from $g(x)$ corresponding to labels $\{y_1, y_2,\cdots, y_n\}$, which are denoted as $g(x)_{\text{top}-n}=\{g_{y_1}(x), g_{y_2}(x), \cdots, g_{y_n}(x)\}$, and the prediction inconsistency is represented by
\begin{equation}
I_{pred} = \Vert f(x)_{\text{top}-n} - g(x)_{\text{top}-n} \Vert_1.
\end{equation}

\textit{Metric 4.} In Metric 4, we use the $\ell_1$ norm of the difference between the whole prediction vectors $f(x)$ and $g(x)$ to measure the prediction inconsistency, which can be described as
\begin{equation}
I_{pred} = \Vert f(x) - g(x) \Vert_1.
\end{equation}
A test sample is identified to be adversarial if the calculated $I_{pred}$ exceeds the pre-defined threshold. Note that although we evaluate all four metrics, Metric 1 is ultimately adopted for implementing the proposed PID, while the results for the remaining three are presented and discussed in Section~\ref{sec:ablation}.

\noindent \textbf{Choice of Auxiliary Models.}
Previous studies have explored improving the transferability of AEs~\cite{wang2021enhancing,lin2019nesterov}, showing that well-crafted AEs can often transfer between naturally trained CNN models. However, transferring AEs from a naturally trained CNN to an adversarially trained CNN remains challenging, which is also verified by our exploration shown in Figure \ref{fig:vgg_adv} and \ref{fig:vgg_at_adv}. This observation suggests that adversarially trained CNN models can serve as effective candidates for the auxiliary model in the proposed PID.

Additionally, Mahmood et al.~\cite{mahmood2021robustness}  reported that AEs generated for CNNs struggle to fool ViT models~\cite{dosovitskiy2021image}, potentially due to architectural differences. This makes ViTs another strong candidate for the auxiliary model, as they can effectively reveal prediction inconsistencies when facing AEs.

Furthermore, pre-trained foundation models trained on large-scale datasets using self-supervised or weakly supervised learning can generalize across multiple downstream tasks~\cite{awais2025foundation}, including the image classification task. Among them, the CLIP model~\cite{radford2021learning}, with its distinct architecture and outstanding zero-shot ability, can also be adopted as the auxiliary model to expose AEs.

Although several options are available for auxiliary models, as observed in Figure \ref{fig:motivation}, ViT-L/16 exhibits the most significant prediction inconsistency from the primal model for AEs.
As a result, we adopt ViT-L/16 as the default auxiliary model in our experiments to demonstrate the effectiveness of our black-box detection method. Results using adversarially trained CNNs and CLIP as auxiliary models are provided and analyzed in Section \ref{sec:ablation}.

\section{Evaluation}
\subsection{Experimental Settings}
\label{sec:setting}
\noindent \textbf{Datasets and Models.} To evaluate the performance of our PID, we utilize CIFAR-10~\cite{krizhevsky2009learning} and ImageNet~\cite{deng2009imagenet}, two representative datasets commonly used in adversarial research.
In the inference stage, we use the whole test set of CIFAR-10, where images that can be correctly classified by primal models are selected and attacked. On ImageNet, we randomly select 1000 correctly classified images from the validation set for each primal model, one from each of the 1000 classes.

On CIFAR-10, we train the VGG19 model~\cite{simonyan2014very} under both natural and adversarial settings. On ImageNet, we use the ResNet50 model~\cite{he2016deep}, and adopt pre-trained versions from both training settings.
We adopt the ViT-L/16 model~\cite{dosovitskiy2021image} pre-trained on ImageNet as the auxiliary model in our PID. When evaluating on CIFAR-10, we fine-tune ViT-L/16 on the same dataset. Details of the primal and auxiliary models, as well as models used for implementing transfer-based black-box attacks, are provided in Section \ref{sec:supp_0}.


\begin{wraptable}{r}{0.5\textwidth}
  \centering
  \caption{Parameters of the implemented attack algorithms.}
  \resizebox{0.48\textwidth}{!}{
    \begin{tabular}{cccc}
    \toprule
    Types & Attack & Norm  & Constraint \\
    \midrule
    White-box & PGD   & $\ell_\infty$ & 1/255, 8/255 \\
    White-box & C\&W & $\ell_2$ & - \\
    White-box & DeepFool & $\ell_2$ & - \\
    Score-based black-box & Square & $\ell_\infty$ & 8/255 \\
    Decision-based black-box & TA    & $\ell_2$ & - \\
    Transfer-based black-box & VNI-FGSM & $\ell_\infty$ & 8/255 or 16/255 \\
    Mixed & AA    & $\ell_\infty$ & 1/255, 8/255 \\
    \bottomrule
    \end{tabular}%
    }
  \label{tab:attacks}%
\end{wraptable}%
\noindent \textbf{Attack Algorithms.} To comprehensively verify the effectiveness of our PID, we adopt white-box, black-box, and mixed adversarial attacks. In particular, to simulate a more challenging adversarial environment, we (1) employ three types of black-box adversarial attacks, given their growing threat in real-world applications, and (2) vary attack parameters to generate perturbations of different magnitudes.

The specific perturbation constraints for three types of attacks are summarized in Table \ref{tab:attacks}. Specifically, for PGD and AA, we set the $\ell_\infty$-constraint to $\epsilon = 1/255$ and $8/255$. For C\&W, we choose $\kappa = 0$ and $1$, where $\kappa$ controls the confidence level of AEs. For VNI-FGSM, we set $\ell_\infty$-constraint to $\epsilon = 8/255$ on CIFAR-10 and $\epsilon = 16/255$ on ImageNet, respectively. The detailed parameters of each implemented attack are given in Section \ref{sec:supp_1}.

\noindent \textbf{Baseline Methods.} We compare our PID with 4 detection methods, namely FS~\cite{xu2017feature}, DiffPure~\cite{nie2022diffusion}, SID~\cite{tian2021detecting}, and EPS-AD~\cite{zhang2023detecting}, where the first two are black-box detection methods, and the last two are white-box detection methods. Each method is combined with the naturally and adversarially trained models, respectively. 
FS and DiffPure serve as baselines on both the CIFAR-10 and ImageNet datasets. It should be noted that DiffPure is originally designed as a defense method. Considering the superior performance of the diffusion model on denoising images~\cite{song2021score,ho2020denoising}, we modify it as a black-box detection method.
SID is adopted as a baseline only on CIFAR-10, as training the dual classifier with the Weighted Average Wavelet Transform layer~\cite{tian2021detecting} on large-scale datasets like ImageNet is computationally expensive. Consistent with the original evaluation~\cite{tian2021detecting}, we do not apply SID to ImageNet. 
Meanwhile, we compare PID with EPS-AD on ImageNet instead. Additionally, these two white-box detection methods will also be evaluated in a black-box detection scenario for a fair comparison with our black-box detection PID. 
We adjust the parameters of all detection methods to obtain their best performance. More detailed information can be found in Section \ref{sec:supp_2}.

\noindent \textbf{Evaluation Metrics.} 
We use the AUC score, the Area Under the Receiver Operating Characteristic curve, to evaluate detection performance. This widely adopted metric provides an aggregate measure across all possible detection thresholds and serves as a unified index in prior works \cite{xu2017feature,zhang2023detecting,tian2021detecting,ma2018characterizing}.
In real-world scenarios, applying PID requires setting a detection threshold. A practical approach is to estimate the distribution of prediction inconsistency values on NEs, and then choose the threshold that corresponds to a fixed False Positive Rate (FPR), such as 5\%.

\subsection{Experimental Results and Analyses}
\begin{table*}[t]
  \centering
  \caption{Comparison of AUC scores (\%) of detecting AEs on CIFAR-10, where NAT (ADV) means the primal model is naturally (adversarially) trained. SID* (in gray) is not directly comparable to others since it is in the ideal, white-box scenario.}
   \resizebox{\textwidth}{!}{
    \begin{tabular}{ccccccccccccc}
    \toprule
         Primal& Detection & PGD   & PGD   & AA    & AA    & C\&W  & C\&W  & \multirow{2}{*}{DeepFool} & \multirow{2}{*}{Square} & \multirow{2}{*}{TA} & \multirow{2}{*}{VNI-FGSM} & \multirow{2}{*}{Average} \\
          Model&     Method   & $\epsilon = \frac{1}{255}$ & $\epsilon = \frac{8}{255}$ & $\epsilon = \frac{1}{255}$ & $\epsilon = \frac{8}{255}$ & $\kappa=0$ & $\kappa=1$ &       &       &       &       &  \\
    \midrule
    \multirow{5}[10]{*}{NAT} & FS    & 83.84  & 56.76  & 87.24  & 63.11  & 93.58  & 89.47  & 92.15  & \underline{92.96}  & 93.10  & 71.42  & 82.36  \\
\cmidrule{2-11}  \cmidrule{12-13}          & DiffPure & \underline{88.93}  & \underline{97.75}  & \underline{91.57}  & \underline{97.90}  &\underline{95.09}  & \underline{95.00}  & \underline{97.02}  & 92.91  & 92.80  & \underline{96.92}  & \underline{94.59}  \\
\cmidrule{2-12}  \cmidrule{13-13}          & SID   & 67.83  & 92.84  & 70.59  & 90.52  & 79.29  & 79.30  & 83.14  & 89.31  & \underline{95.36}  & 73.03  & 82.12  \\
\cmidrule{2-13}          &\cellcolor{gray!30} SID*  &\cellcolor{gray!30} 68.09  &\cellcolor{gray!30} 99.35  &\cellcolor{gray!30} 72.23  &\cellcolor{gray!30} 99.23  &\cellcolor{gray!30} 85.54  &\cellcolor{gray!30} 86.21  &\cellcolor{gray!30} 86.83  &\cellcolor{gray!30} 94.98  &\cellcolor{gray!30} 97.65  &\cellcolor{gray!30} 85.65  &\cellcolor{gray!30} 87.58  \\
\cmidrule{2-13}          & PID (Ours)   & \textbf{99.81 } & \textbf{98.54 } & \textbf{99.85 } & \textbf{98.71 } & \textbf{99.93 } & \textbf{99.45 } & \textbf{99.86 } & \textbf{99.88 } & \textbf{99.85 } & \textbf{97.02 } & \textbf{99.29 } \\
    \midrule
    \multirow{5}[10]{*}{ADV} & FS    & 49.40  & 58.22  & 49.29  & 61.67  & 81.79  & 84.22  & 70.81  & 55.29  & 76.40  & 50.88  & 63.80  \\
\cmidrule{2-13}          & DiffPure & 50.81  & 74.55  & \underline{50.77}  & 76.38  & 87.00  & \underline{88.55}  & 83.77  & 63.60  & 84.38  & \underline{52.77}  & 71.26  \\
\cmidrule{2-13}          & SID   & \underline{51.96}  & \underline{75.90}  & 40.09  & \underline{78.08}  & \underline{89.92}  & 87.07  & \underline{89.64}  & \underline{82.80}  & \underline{90.42}  & 52.49  & \underline{73.84}  \\
\cmidrule{2-13}          &\cellcolor{gray!30} SID*  &\cellcolor{gray!30} 63.15  &\cellcolor{gray!30} 75.69  &\cellcolor{gray!30} 51.67  &\cellcolor{gray!30} 76.25  &\cellcolor{gray!30} 90.10  &\cellcolor{gray!30} 89.27  &\cellcolor{gray!30} 92.28  &\cellcolor{gray!30} 80.80  & \cellcolor{gray!30}93.52  &\cellcolor{gray!30} 62.88  &\cellcolor{gray!30} 77.56  \\
\cmidrule{2-13}          & PID (Ours)   & \textbf{99.87 } & \textbf{99.68 } & \textbf{99.88 } & \textbf{99.69 } & \textbf{99.91 } & \textbf{99.90 } & \textbf{99.55 } & \textbf{99.82 } & \textbf{99.82 } & \textbf{94.90 } &  \textbf{99.30 } \\ 
    \bottomrule
    \end{tabular}%
    }
  \label{tab:cifar10}%
\end{table*}%

\noindent \textbf{Results on CIFAR-10.} Evaluation of detection methods is summarized in Table \ref{tab:cifar10}.
As described in Section \ref{sec:design}, our proposed PID is a black-box detection method, which utilizes the outputs of primal models without access to their internal outputs or parameters. For a fair comparison, SID is first performed in a black-box way. To be specific, SID is trained using the AEs against naturally and adversarially trained VGG16 models and evaluated on the AEs against naturally and adversarially trained VGG19 models, respectively. Meanwhile, the SID in the white-box detection scenario is also implemented (denoted as SID* in Table \ref{tab:cifar10}).

When working with the naturally trained VGG19 model, PID consistently achieves high AUC scores, with an average AUC score of 99.29\% across 10 attacks. The performance of FS degrades dramatically when detecting strong attacks, e.g., its AUC scores are 56.76\% and 63.11\% when detecting PGD-8/255 and AA-8/255 attacks, respectively, which results from that strong AEs are actually more robust than NEs after being squeezed. DiffPure also achieves a good detection performance with an average AUC score of 94.59\%, demonstrating its effectiveness on denoising AEs. 
For SID in the black-box scenario, its performance is limited, especially when detecting small-scale AEs such as PGD-1/255 and AA-1/255. After performing SID in the white-box scenario, where the detector is trained and tested on AEs generated by the same attack against the same primal model, its performance improves, reaching an average AUC score of 87.58\%. Nevertheless, this score remains constrained.

After being integrated with the adversarially trained model, other detection methods exhibit a decline in performance, which is reflected by the fact that their average AUC scores are less than 80\%. 
In contrast, our PID still achieves significantly higher AUC performance compared to other methods, exhibiting its good generalization and flexibility.
This performance gap can be explained by two key factors. First, adversarial training makes the model less overconfident \cite{grabinski2022robust}, leading to less noticeable differences in predictions before and after image transformations or denoising, which weakens methods that rely on such differences. Second, the increased robustness of adversarially trained models leads to fewer successful AEs, resulting in insufficient training samples for the detector.

\begin{table*}[t]
  \centering
  \caption{Comparison of AUC scores (\%) of detecting AEs on ImageNet, where NAT (ADV) means the primal model is naturally (adversarially) trained, EPS-AD* (in gray) is not directly comparable to others since it is in the ideal, white-box scenario.}
   \resizebox{\textwidth}{!}{
    \begin{tabular}{ccccccccccccc}
    \toprule
          Primal& Detection  & PGD   & PGD   & AA    & AA    & C\&W  & C\&W  & \multirow{2}[2]{*}{DeepFool} & \multirow{2}[2]{*}{Square} & \multirow{2}[2]{*}{TA} & \multicolumn{1}{c}{\multirow{2}[2]{*}{VNI-FGSM}} & \multicolumn{1}{c}{\multirow{2}[2]{*}{Average}} \\
          Model & Method & $\epsilon = \frac{1}{255}$ & $\epsilon = \frac{8}{255}$ & $\epsilon = \frac{1}{255}$ & $\epsilon = \frac{8}{255}$ & $\kappa=0$ & $\kappa=1$ &       &       &       &       &  \\
    \midrule
    \multirow{5}[10]{*}{NAT} & FS    & 91.50  & 24.48  & 90.07  & 32.08  & \underline{92.46}  & 84.79  & \underline{95.43}  & \underline{87.64}  & \underline{88.17}  & 74.75  & 76.14  \\
\cmidrule{2-13}          & DiffPure & 96.38  & 97.59  & \underline{97.70}  & 97.97  & 91.53  & 95.42  & 86.55  & 85.22  & 83.62  & 96.30  & \underline{92.83}  \\
\cmidrule{2-13}          & EPS-AD & \underline{96.64}  & \textbf{99.89 } & 96.90  & \textbf{99.77 } & 92.07  & \textbf{99.84 } & 55.48  & 62.43  & 55.55  & \textbf{99.38 } & 85.80  \\
\cmidrule{2-13}          & EPS-AD* &\cellcolor{gray!30} 99.33  &\cellcolor{gray!30} 99.91  &\cellcolor{gray!30} 99.55  &\cellcolor{gray!30} 99.88  &\cellcolor{gray!30} 94.85  &\cellcolor{gray!30} 99.95  &\cellcolor{gray!30} 57.55  &\cellcolor{gray!30} 65.24  &\cellcolor{gray!30} 58.68  &\cellcolor{gray!30} 99.95  &\cellcolor{gray!30} 87.49  \\
\cmidrule{2-13}          & PID (Ours)   & \textbf{98.45 } & \underline{98.54}  & \textbf{98.74 } & \underline{98.90}  & \textbf{98.17 } & \underline{98.20}  & \textbf{98.36 } & \textbf{97.80 } & \textbf{97.80 } & \underline{98.11}  & \textbf{98.31 } \\
    \midrule
    \multirow{5}[10]{*}{ADV} & FS    & 48.86  & 72.00  & 49.04  & 75.59  & 72.07  & 76.89  & 74.75  & 59.99  & 81.39  & 62.77  & 67.34  \\
\cmidrule{2-13}          & DiffPure & 47.33  & 82.71  & 50.26  & 86.91  & 56.96  & 72.16  & 78.09  & 58.32  & 79.41  & 62.18  & 67.43  \\
\cmidrule{2-13}          & EPS-AD & \underline{87.56}  & \textbf{99.83 } & \underline{85.48}  & \textbf{99.89 } & \underline{96.58}  & \textbf{99.82 } & \underline{94.36}  & \underline{66.03}  & \underline{85.76}  & \textbf{100.00 } & \underline{91.53}  \\
\cmidrule{2-13}          & EPS-AD* &\cellcolor{gray!30} 95.20  &\cellcolor{gray!30} 99.12  &\cellcolor{gray!30} 95.27  &\cellcolor{gray!30} 99.14  &\cellcolor{gray!30} 97.78  &\cellcolor{gray!30} 99.65  &\cellcolor{gray!30} 95.52  &\cellcolor{gray!30} 65.35  &\cellcolor{gray!30} 88.60  &\cellcolor{gray!30} 100.00  &\cellcolor{gray!30} 93.56  \\
\cmidrule{2-13}          & PID (Ours)   & \textbf{95.98} & \underline{96.17}  & \textbf{98.20} & \underline{97.83}  & \textbf{97.23} & \underline{97.11}  & \textbf{98.06} & \textbf{96.90} & \textbf{98.34} & \underline{92.23}  & \textbf{96.81} \\
    \bottomrule
    \end{tabular}%
    }
  \label{tab:imagenet}%
\end{table*}%

\noindent \textbf{Results on ImageNet.} The detection performance of our PID and the other three baselines on ImageNet is shown in Table \ref{tab:imagenet}. Following~\cite{zhang2023detecting}, EPS-AD in the white-box detection scenario is trained using the AEs generated by $\ell_\infty$-FGSM and $\ell_2$-FGSM attacks with a perturbation size of 1/255 against the primal ResNet50 models (denoted as EPS-AD* in Table \ref{tab:imagenet}). In the black-box detection scenario, the FGSM AEs used to train the EPS-AD are crafted against the naturally and adversarially trained ResNet101 models. It is worth noting that this setup is relatively mild, as the primal models employed in the training and evaluation of the detector share very similar architectures. 

It can be seen from Table \ref{tab:imagenet} that, our PID still outperforms other detection methods when combined with both naturally and adversarially trained models, where it consistently achieves either the best or the second-best detection performance against each attack, demonstrating its effectiveness in detecting AEs. 
Similar to the experimental results shown in Table \ref{tab:cifar10}, the detection performance of FS and DiffPure exhibits a noticeable decrease after working with the adversarially trained model.
While EPS-AD exhibits strong performance in detecting AEs across multiple attack types, its effectiveness drops significantly when faced with soft-label and hard-label black-box attacks, i.e., Square and TA attacks, either combined with naturally or adversarially trained models. This also emphasizes the necessity of introducing black-box attacks in the reliable evaluation of detection methods.

\subsection{Adaptive Attack}
We now evaluate the robustness of our PID under the adaptive attack, where the attacker has access to the full knowledge of both primal and auxiliary models. Specifically, the aim of the attacker is to mislead $f(x)$ and $g(x)$ simultaneously. Since in Metric 1, $I_{pred}=1-g_y(x)$ is used to measure the prediction inconsistency, an intuitive adaptive attack strategy is to maximize $g_y(x)$. This can be achieved by performing a targeted attack against two models, forcing both the primal model and the auxiliary model to misclassify the AE into the same label, which can be expressed as,
\begin{equation}
\min_r \mathcal{L}(f(x+r),t) + \lambda\mathcal{L}(g(x+r),t),\, \text{s.t.} \, \Vert r\Vert_\infty\leq \epsilon,
\label{eq:adaptive}
\end{equation}
where $t$ is the targeted label satisfying $t\neq y_{true}$, $y_{true}$ is the ground-truth label of $x$, $\mathcal{L}(\cdot,\cdot)$ is the loss function, $\epsilon$ is the perturbation constraint of the perturbation $r$, and $\lambda$ is a trade-off parameter.

The performance of PID against the adaptive attack on CIFAR-10 and ImageNet is shown in Table \ref{tab:adaptive}. We set $\lambda=1$ and use the PGD attack strategy to find the perturbation $r$, where the perturbation constraint $\epsilon=8/255$. 
As shown in Table \ref{tab:adaptive}, PID maintains strong detection performance when it works with the adversarially trained model. 
However, its performance declines when combined with the naturally trained model. This can be attributed to the vulnerability of both CNNs and ViTs to adversarial threats~\cite{mahmood2021robustness}, allowing the adaptive attack to jointly mislead both models.

It is important to note that adaptive attack represents an idealized scenario, where the attacker has complete knowledge of the original classifier and the detection method, despite the fact that our PID only has access to the original classifier's outputs. Many detection methods have been broken under this condition~\cite{bryniarski2021evading}. On the other hand, our results suggest that pairing detection with a robust model, such as the adversarially trained model, can mitigate even extremely strong adaptive attacks. This emphasizes the value of generalizable detection techniques that can be applied across diverse model settings, including both standard and defense-enhanced models, which still remains to be explored.

\begin{table}[!t]
  \centering
    \begin{minipage}[!t]{0.48\textwidth}
      \centering
      \captionof{table}{Detection performance of AUC scores (\%) on PGD vs. an adaptive attack.}
       \resizebox{0.95\columnwidth}{!}{
        \begin{tabular}{cccc}
        \toprule
         Dataset     & Primal Model      & PGD&Adaptive \\
        \midrule
        \multirow{2}[2]{*}{CIFAR-10} & NAT  &98.54 &58.13  \\
              & ADV   & 99.68 &87.19  \\
        \midrule
        \multirow{2}[2]{*}{ImageNet} & NAT   &98.54 &79.28  \\
              & ADV   &96.17 &86.00  \\
        \bottomrule
        \end{tabular}%
        }
      \label{tab:adaptive}%
    \end{minipage}
    \hfill
    \begin{minipage}[!t]{0.48\textwidth}
      \centering
      \captionof{table}{Detection performance of PID by employing different metrics for quantifying prediction inconsistency on CIFAR-10. The average AUC scores (\%) over 10 attacks are shown, with full results available in Section \ref{sec:supp_3}.}
      \resizebox{0.95\columnwidth}{!}{
        \begin{tabular}{ccc}
        \toprule
        Metric & NAT   & ADV \\
        \midrule
        $I_{pred} =-g_y(x)$  & \textbf{99.29 } & \textbf{99.30 } \\
        $I_{pred} =f_y(x)-g_y(x)$ & 98.58  & \underline{96.74}  \\
        $I_{pred} = \Vert f(x)_{\text{top}-n} - g(x)_{\text{top}-n} \Vert_1$ & 98.75  & 77.76  \\
        $I_{pred} = \Vert f(x) - g(x) \Vert_1$ & \underline{98.96}  & 78.94  \\
        \bottomrule
        \end{tabular}%
        }
      \label{tab:sotf_label}%
    \end{minipage}
\end{table}

\subsection{Ablation Study}
\label{sec:ablation}
\noindent \textbf{Metrics for Quantifying Prediction Inconsistency.} Here, we discuss how the metrics for quantifying prediction inconsistency impact the effectiveness of the PID.
Four metrics described in Section \ref{sec:design} are employed to measure prediction inconsistency for implementing PID, where we use $n=3$ in Metric 3. The experimental results on CIFAR-10 are summarized in Table \ref{tab:sotf_label}.
We can observe that four metrics yield similar results when the protected primal model is naturally trained, yet Metric 1 remains the best. However, after the primal model is adversarially trained, it becomes less overconfident toward both NEs and AEs. As a result, the differences between 
confidence scores from $f(x)$ and $g(x)$ tend to be larger for NEs and smaller for AEs—particularly for weak AEs with low confidence scores on incorrectly predicted labels. This reduction in disparity makes it more challenging to tell NEs and AEs apart. While Metric 1 only requires the label from $f(x)$, avoiding the fluctuation of confidence scores brought by the training strategy, making the PID consistently effective.

\begin{wraptable}{r}{0.5\textwidth}
  \centering
  \caption{Detection performance of PID by employing different models as the auxiliary model on CIFAR-10 and ImageNet, where NAT (ADV) means the primal model is naturally (adversarially) trained. The average AUC scores (\%) over 10 attacks are shown, with full results available in Section \ref{sec:supp_4}.}
  \resizebox{0.45\textwidth}{!}{
    \begin{tabular}{cccc}
    \toprule
    Dataset & Auxiliary Model & NAT   & ADV \\
    \midrule
    \multirow{3}[6]{*}{CIFAR-10} & ResNet34\_ADV & 95.14  & 92.75  \\
\cmidrule{2-4}          & ViT-L/16 & \textbf{99.29 } & \textbf{99.30 } \\
\cmidrule{2-4}          & CLIP-ViT-L/14 & \underline{98.36}  & \underline{98.37}  \\
    \midrule
    \multirow{3}[6]{*}{ImageNet} & ConvNeXt-S\_ADV & \underline{96.93}  & \underline{95.06}  \\
\cmidrule{2-4}          & ViT-L/16 & \textbf{98.31 } & \textbf{96.81 } \\
\cmidrule{2-4}          & CLIP-ViT-L/14 & 95.34  & 94.28  \\
    \bottomrule
    \end{tabular}%
    }
  \label{tab:three_model}%
\end{wraptable}
\noindent \textbf{Choice of Auxiliary Model.} 
As discussed in Section \ref{sec:design}, in addition to ViT, the adversarially trained CNN
and CLIP model can also serve as the auxiliary model in our PID, and corresponding experimental results are summarized in Table \ref{tab:three_model}.
On CIFAR-10, we additionally use the adversarially trained ResNet34 (denoted as ResNet34\_ADV) and CLIP-ViT-L/14. We can observe that CLIP-ViT-L/14 can also achieve good detection performance when combined with naturally and adversarially trained models. 
The performance of ResNet34\_ADV degrades slightly when working with the VGG19\_ADV, as its prediction inconsistency with the primal model on AEs is less pronounced, which can also be observed in Figure \ref{fig:vgg_at_adv}.
On ImageNet, we employ the adversarially pre-trained ConvNeXt-S~\cite{liu2022convnet} (denoted as ConvNeXt-S\_ADV) and CLIP-ViT-L/14. The better detection performance of ConvNeXt-S\_ADV compared to CLIP is likely due to its higher accuracy on clean images, yielding lower $I_{pred}$ on NEs and making it easier to distinguish AEs from NEs.

Across both datasets, ViT-L/16 consistently achieves the highest AUC scores, regardless of whether it works with the naturally or adversarially trained models, which can be attributed to its high clean accuracy and significant architecture difference from CNNs. Additionally, although the detection performance of the adversarially trained CNN and CLIP models varies across datasets, their average AUC scores still surpass those of the other four baselines shown in Table \ref{tab:cifar10} and \ref{tab:imagenet}, confirming the effectiveness and the flexibility of the proposed PID. Although we use these three types of models as auxiliary models in our PID, we believe that the choice of the auxiliary model is not limited to these and needs further exploration.

\section{Conclusion}
This paper proposes a lightweight detection method named PID, which leverages the prediction inconsistency between the primal and auxiliary models to effectively detect AEs without requiring any prior model-specific knowledge.
Our method demonstrates strong generalization, as it not only maintains compatibility with both naturally and adversarially trained models, but also achieves consistently high detection performance across a comprehensive evaluation, including white-box, black-box, and mixed attacks with varying perturbation strengths on two widely used datasets.

\bibliographystyle{unsrt}
\bibliography{main}

\begin{thebibliography}{10}

\bibitem{goodfellow2014explaining}
Ian~J Goodfellow, Jonathon Shlens, and Christian Szegedy.
\newblock Explaining and harnessing adversarial examples.
\newblock In {\em Proceedings of the International Conference on Learning Representations}, 2015.

\bibitem{carlini2017towards}
Nicholas Carlini and David Wagner.
\newblock Towards evaluating the robustness of neural networks.
\newblock In {\em 2017 ieee symposium on security and privacy (sp)}, pages 39--57. IEEE, 2017.

\bibitem{ma2021understanding}
Xingjun Ma, Yuhao Niu, Lin Gu, Yisen Wang, Yitian Zhao, James Bailey, and Feng Lu.
\newblock Understanding adversarial attacks on deep learning based medical image analysis systems.
\newblock {\em Pattern Recognition}, 110:107332, 2021.

\bibitem{badjie2024adversarial}
Bakary Badjie, Jos{\'e} Cec{\'\i}lio, and Antonio Casimiro.
\newblock Adversarial attacks and countermeasures on image classification-based deep learning models in autonomous driving systems: A systematic review.
\newblock {\em ACM Computing Surveys}, 57(1):1--52, 2024.

\bibitem{madry2017towards}
Aleksander Madry, Aleksandar Makelov, Ludwig Schmidt, Dimitris Tsipras, and Adrian Vladu.
\newblock Towards deep learning models resistant to adversarial attacks.
\newblock In {\em Proceedings of the International Conference on Learning Representations}, 2018.

\bibitem{wong2020fast}
Eric Wong, Leslie Rice, and J~Zico Kolter.
\newblock Fast is better than free: Revisiting adversarial training.
\newblock In {\em International Conference on Learning Representations}, 2020.

\bibitem{liu2024comprehensive}
Chang Liu, Yinpeng Dong, Wenzhao Xiang, Xiao Yang, Hang Su, Jun Zhu, Yuefeng Chen, Yuan He, Hui Xue, and Shibao Zheng.
\newblock A comprehensive study on robustness of image classification models: Benchmarking and rethinking.
\newblock {\em International Journal of Computer Vision}, pages 1--23, 2024.

\bibitem{xu2017feature}
Weilin Xu, David Evans, and Yanjun Qi.
\newblock Feature squeezing: Detecting adversarial examples in deep neural networks.
\newblock In {\em Proceedings of the 25th Network and Distributed System Security Symposium (NDSS)}, 2018.

\bibitem{zhang2023detecting}
Shuhai Zhang, Feng Liu, Jiahao Yang, Yifan Yang, Changsheng Li, Bo~Han, and Mingkui Tan.
\newblock Detecting adversarial data by probing multiple perturbations using expected perturbation score.
\newblock In {\em International Conference on Machine Learning}, pages 41429--41451. PMLR, 2023.

\bibitem{tian2021detecting}
Jinyu Tian, Jiantao Zhou, Yuanman Li, and Jia Duan.
\newblock Detecting adversarial examples from sensitivity inconsistency of spatial-transform domain.
\newblock In {\em Proceedings of the AAAI Conference on Artificial Intelligence}, volume~35, pages 9877--9885, 2021.

\bibitem{ma2018characterizing}
Xingjun Ma, Bo~Li, Yisen Wang, Sarah~M Erfani, Sudanthi Wijewickrema, Grant Schoenebeck, Dawn Song, Michael~E Houle, and James Bailey.
\newblock Characterizing adversarial subspaces using local intrinsic dimensionality.
\newblock In {\em International Conference on Learning Representations}, 2018.

\bibitem{wang2023detecting}
Qian Wang, Yongqin Xian, Hefei Ling, Jinyuan Zhang, Xiaorui Lin, Ping Li, Jiazhong Chen, and Ning Yu.
\newblock Detecting adversarial faces using only real face self-perturbations.
\newblock In {\em Proceedings of the Thirty-Second International Joint Conference on Artificial Intelligence}, pages 1488--1496, 2023.

\bibitem{tian2018detecting}
Shixin Tian, Guolei Yang, and Ying Cai.
\newblock Detecting adversarial examples through image transformation.
\newblock In {\em Proceedings of the AAAI conference on artificial intelligence}, volume~32, 2018.

\bibitem{aldahdooh2022adversarial}
Ahmed Aldahdooh, Wassim Hamidouche, Sid~Ahmed Fezza, and Olivier D{\'e}forges.
\newblock Adversarial example detection for dnn models: A review and experimental comparison.
\newblock {\em Artificial Intelligence Review}, pages 1--60, 2022.

\bibitem{croce2020reliable}
Francesco Croce and Matthias Hein.
\newblock Reliable evaluation of adversarial robustness with an ensemble of diverse parameter-free attacks.
\newblock In {\em International Conference on Machine Learning}, pages 2206--2216. PMLR, 2020.

\bibitem{croce2020minimally}
Francesco Croce and Matthias Hein.
\newblock Minimally distorted adversarial examples with a fast adaptive boundary attack.
\newblock In {\em International Conference on Machine Learning}, pages 2196--2205. PMLR, 2020.

\bibitem{andriushchenko2020square}
Maksym Andriushchenko, Francesco Croce, Nicolas Flammarion, and Matthias Hein.
\newblock Square attack: a query-efficient black-box adversarial attack via random search.
\newblock In {\em European Conference on Computer Vision}, pages 484--501. Springer, 2020.

\bibitem{kingma2014adam}
Diederik~P Kingma and Jimmy Ba.
\newblock Adam: A method for stochastic optimization.
\newblock In {\em International Conference on Learning Representations}, 2015.

\bibitem{moosavi2016deepfool}
Seyed-Mohsen Moosavi-Dezfooli, Alhussein Fawzi, and Pascal Frossard.
\newblock Deepfool: a simple and accurate method to fool deep neural networks.
\newblock In {\em Proceedings of the IEEE Conference on Computer Cision and Pattern Recognition}, pages 2574--2582, 2016.

\bibitem{wang2022triangle}
Xiaosen Wang, Zeliang Zhang, Kangheng Tong, Dihong Gong, Kun He, Zhifeng Li, and Wei Liu.
\newblock Triangle attack: A query-efficient decision-based adversarial attack.
\newblock In {\em European Conference on Computer Vision}, pages 156--174. Springer, 2022.

\bibitem{wang2021enhancing}
Xiaosen Wang and Kun He.
\newblock Enhancing the transferability of adversarial attacks through variance tuning.
\newblock In {\em Proceedings of the IEEE/CVF conference on computer vision and pattern recognition}, pages 1924--1933, 2021.

\bibitem{lin2019nesterov}
Jiadong Lin, Chuanbiao Song, Kun He, Liwei Wang, and John~E Hopcroft.
\newblock Nesterov accelerated gradient and scale invariance for adversarial attacks.
\newblock In {\em International Conference on Learning Representations}, 2019.

\bibitem{song2021score}
Yang Song, Jascha Sohl-Dickstein, Diederik~P Kingma, Abhishek Kumar, Stefano Ermon, and Ben Poole.
\newblock Score-based generative modeling through stochastic differential equations.
\newblock In {\em International Conference on Learning Representations}, 2021.

\bibitem{dosovitskiy2021image}
Alexey Dosovitskiy, Lucas Beyer, Alexander Kolesnikov, Dirk Weissenborn, Xiaohua Zhai, Thomas Unterthiner, Mostafa Dehghani, Matthias Minderer, Georg Heigold, Sylvain Gelly, et~al.
\newblock An image is worth 16x16 words: Transformers for image recognition at scale.
\newblock In {\em International Conference on Learning Representations}, 2021.

\bibitem{radford2021learning}
Alec Radford, Jong~Wook Kim, Chris Hallacy, Aditya Ramesh, Gabriel Goh, Sandhini Agarwal, Girish Sastry, Amanda Askell, Pamela Mishkin, Jack Clark, et~al.
\newblock Learning transferable visual models from natural language supervision.
\newblock In {\em International conference on machine learning}, pages 8748--8763. PmLR, 2021.

\bibitem{jia2024improving}
Xiaojun Jia, Yong Zhang, Xingxing Wei, Baoyuan Wu, Ke~Ma, Jue Wang, and Xiaochun Cao.
\newblock Improving fast adversarial training with prior-guided knowledge.
\newblock {\em IEEE Transactions on Pattern Analysis and Machine Intelligence}, 2024.

\bibitem{mahmood2021robustness}
Kaleel Mahmood, Rigel Mahmood, and Marten Van~Dijk.
\newblock On the robustness of vision transformers to adversarial examples.
\newblock In {\em Proceedings of the IEEE/CVF international conference on computer vision}, pages 7838--7847, 2021.

\bibitem{awais2025foundation}
Muhammad Awais, Muzammal Naseer, Salman Khan, Rao~Muhammad Anwer, Hisham Cholakkal, Mubarak Shah, Ming-Hsuan Yang, and Fahad~Shahbaz Khan.
\newblock Foundation models defining a new era in vision: a survey and outlook.
\newblock {\em IEEE Transactions on Pattern Analysis and Machine Intelligence}, 2025.

\bibitem{krizhevsky2009learning}
Alex Krizhevsky, Geoffrey Hinton, et~al.
\newblock Learning multiple layers of features from tiny images.
\newblock 2009.

\bibitem{deng2009imagenet}
Jia Deng, Wei Dong, Richard Socher, Li-Jia Li, Kai Li, and Li~Fei-Fei.
\newblock Imagenet: A large-scale hierarchical image database.
\newblock In {\em 2009 IEEE Conference on Computer Vision and Pattern Recognition}, pages 248--255. Ieee, 2009.

\bibitem{simonyan2014very}
Karen Simonyan and Andrew Zisserman.
\newblock Very deep convolutional networks for large-scale image recognition.
\newblock In {\em International Conference on Learning Representations}, 2015.

\bibitem{he2016deep}
Kaiming He, Xiangyu Zhang, Shaoqing Ren, and Jian Sun.
\newblock Deep residual learning for image recognition.
\newblock In {\em Proceedings of the IEEE conference on computer vision and pattern recognition}, pages 770--778, 2016.

\bibitem{nie2022diffusion}
Weili Nie, Brandon Guo, Yujia Huang, Chaowei Xiao, Arash Vahdat, and Animashree Anandkumar.
\newblock Diffusion models for adversarial purification.
\newblock In {\em International Conference on Machine Learning}, pages 16805--16827. PMLR, 2022.

\bibitem{ho2020denoising}
Jonathan Ho, Ajay Jain, and Pieter Abbeel.
\newblock Denoising diffusion probabilistic models.
\newblock {\em Advances in Neural Information Processing Systems}, 33:6840--6851, 2020.

\bibitem{grabinski2022robust}
Julia Grabinski, Paul Gavrikov, Janis Keuper, and Margret Keuper.
\newblock Robust models are less over-confident.
\newblock {\em Advances in Neural Information Processing Systems}, 35:39059--39075, 2022.

\bibitem{bryniarski2021evading}
Oliver Bryniarski, Nabeel Hingun, Pedro Pachuca, Vincent Wang, and Nicholas Carlini.
\newblock Evading adversarial example detection defenses with orthogonal projected gradient descent.
\newblock In {\em International Conference on Learning Representations}, 2021.

\bibitem{liu2022convnet}
Zhuang Liu, Hanzi Mao, Chao-Yuan Wu, Christoph Feichtenhofer, Trevor Darrell, and Saining Xie.
\newblock A convnet for the 2020s.
\newblock In {\em Proceedings of the IEEE/CVF conference on computer vision and pattern recognition}, pages 11976--11986, 2022.

\bibitem{kim2020torchattacks}
Hoki Kim.
\newblock Torchattacks: A pytorch repository for adversarial attacks.
\newblock {\em arXiv preprint arXiv:2010.01950}, 2020.

\bibitem{dong2020benchmarking}
Yinpeng Dong, Qi-An Fu, Xiao Yang, Tianyu Pang, Hang Su, Zihao Xiao, and Jun Zhu.
\newblock Benchmarking adversarial robustness on image classification.
\newblock In {\em proceedings of the IEEE/CVF conference on computer vision and pattern recognition}, pages 321--331, 2020.

\bibitem{dhariwal2021diffusion}
Prafulla Dhariwal and Alexander Nichol.
\newblock Diffusion models beat gans on image synthesis.
\newblock {\em Advances in Neural Information Processing Systems}, 34:8780--8794, 2021.

\end{thebibliography}

\newpage
\appendix
\section{Appendix}

\subsection{Limitations}
\label{sec:limitation}
We observe a prediction inconsistency between primal and auxiliary models on AEs, which is analyzed in Section \ref{sec:motivation}. This phenomenon inspires the design of our detection method, contributing to both its effectiveness and efficiency. Nonetheless, a more rigorous theoretical understanding, such as establishing upper and lower bounds of this inconsistency, remains an open direction for future work.
Additionally, the auxiliary model used in this work requires training, which may limit the applicability of the method in highly restrictive scenarios where training data is unavailable.
However, our experimental results show that the method remains effective even when using readily accessible pre-trained foundation models, such as CLIP, mitigating the limitation in scenarios where training data is unavailable.
Although CLIP is not the best-performing auxiliary model in our experiments, it consistently outperforms other baselines on both datasets, making it a competitive alternative in such settings.
Finally, under adaptive attacks, where the attacker has full knowledge of both the protected model and the detector, our method shows a performance drop. While evaluating detection methods under such idealized conditions is valuable, we believe it is equally important to develop detection methods that are compatible with both naturally trained models and those equipped with defenses such as adversarial training. Such robustness can provide broader protection and help prevent total failure of the detection system in real-world applications.

\subsection{Broader Impact}
\label{sec:impact}
This work develops a lightweight method for detecting adversarial examples and includes both white-box and more realistic black-box attack settings in the evaluation. Since black-box attacks better reflect real-world threat scenarios, where attackers typically lack full access to model details, this work calls for greater attention to evaluating detection under such settings. Socially, the research contributes to AI security by efficiently detecting adversarial examples, providing a scheme for effectively protecting AI systems in safety-critical applications. However, it may also motivate the development of more advanced and evasive attacks.

\subsection{Details on Employed Models}
\label{sec:supp_0}
\textbf{Primal Models.} On CIFAR-10, we adopt VGG19 as the primal model. Under natural training, the model is trained for 150 epochs with a batch size of 128 using the SGD optimizer, with a weight decay of $5 \times 10^{-4}$. The initial learning rate is 0.1, which is reduced by a factor of 0.1 at the 50th and 100th epochs. It achieves 92.21\% accuracy on the CIFAR-10 test set. For adversarial training, we follow \cite{madry2017towards} and use PGD with an $\ell_\infty$ perturbation bound of 8/255, step size of 2/255, and 10 iterations. The model is adversarially trained for 200 epochs with a batch size of 128, starting from an initial learning rate of 0.005, decayed by 0.1 at the 50th and 100th epochs. It achieves 74.19\% accuracy on the CIFAR-10 test set. 
On ImageNet, we adopt ResNet50 as the primal model and adopt pre-trained weights under both natural\footnote{\label{fn:a}\url{https://github.com/pytorch/vision/blob/main/torchvision/models/resnet.py}} and adversarial\footnote{\label{fn:b}\url{https://github.com/thu-ml/ares/tree/main/robust_training}} training settings.

\textbf{Auxiliary Models.} In this work, the employed auxiliary models fall into three categories: adversarially trained CNNs, ViTs, and CLIP models.  
For adversarially trained CNNs, we adopt the adversarially trained ResNet34 (denoted as ResNet34\_ADV) on CIFAR-10, which is trained under the same setting as the adversarially trained VGG19. It achieves 83.44\% accuracy on the CIFAR-10 test set. On ImageNet, we adopt the pre-trained ConvNeXt-S under the adversarial training settings\textsuperscript{\ref{fn:b}} (denoted as ConvNeXt-S\_ADV).
For ViTs, we use the pre-trained ViT-L/16\footnote{\url{https://github.com/pytorch/vision/blob/main/torchvision/models/vision_transformer.py}} on ImageNet. On CIFAR-10, we fine-tune it for 20 epochs with a learning rate of $1 \times 10^{-5}$, achieving 98.35\% test accuracy.  
For CLIP models, we use the pre-trained CLIP-ViT-L/14\footnote{\url{https://huggingface.co/openai/clip-vit-large-patch14}} on both CIFAR-10 and ImageNet.

\textbf{Models in Transfer-based Black-box Attacks.} To implement the VNI-FGSM attack on both CIFAR-10 and ImageNet, we train substitute models with architectures similar to the corresponding primal models to enhance adversarial transferability. On CIFAR-10, we use the naturally trained VGG13, which is trained under the same setting as the naturally trained VGG19, achieving 92.84\% test accuracy. On ImageNet, we adopt the pre-trained ResNet152\textsuperscript{\ref{fn:a}}.

\subsection{Details on Implemented Attacks}
\label{sec:supp_1}
\textbf{White-box attacks.} The $\ell_\infty$-constraints of \textbf{PGD} attack are set to 1/255 and 8/255. The former uses a step size of 1/255 with 2 iterations, while the latter uses a step size of 2/255 with 10 iterations.
For \textbf{C\&W} attack, the hyper-parameter $\kappa$ that controls the confidence level of the generated AE is set to be 0 and 1 to generate AEs with different strengths. 
For \textbf{DeepFool} attack, the maximum numbers of iterations are 30 and 500 on CIFAR-10 and ImageNet, respectively, and the constant used to enlarge the last step to cross over the decision boundary is 0.02.

\noindent \textbf{Black-box attacks.} Following the implementation of \textbf{TA} in~\cite{wang2022triangle}, the maximum number of iterations in each subspace $N=2$, and the dimension of directional line $d=3$. For updating angle $\alpha$, the change rate $\gamma=0.01$, the constant $\lambda = 0.05$, and the parameter restricting the upper and lower bounds $\tau = 0.1$ are used. The query numbers are set to be 500 and 1000 on CIFAR-10 and ImageNet, respectively.
The constraint of \textbf{Square} attack is set to 8/255, and the query limits are set as follows: 5000 for the naturally trained model on CIFAR-10, 10,000 for the adversarially trained model on CIFAR-10, 10,000 for the naturally trained model on ImageNet, and 20,000 for the adversarially trained model on ImageNet.
For the transfer-based \textbf{VNI-FGSM} attack, VGG13 and ResNet152 models are used as substitute models on CIFAR-10 and ImageNet, respectively. Following~\cite{wang2021enhancing}, the number of sampled examples in the neighborhood $N=20$, the upper bound of neighborhood $\beta=1.5$, and the decay factor $\mu = 1.0$. The perturbation sizes are set to $\epsilon=8/255$ and $\epsilon=16/255$ on CIFAR-10 and ImageNet, respectively, the number of iteration $n=10$, and step size $\alpha = \epsilon/n$.

\noindent \textbf{Mixed attacks.} \textbf{AA} is an ensemble attack consisting of the untargeted APGD-CE, targeted APGD-DLR, targeted FAB, and the untargeted Square attack with 5000 query times. The $\ell_\infty$-constraints of AA are set to be 1/255 and 8/255 for each model on CIFAR-10 and ImageNet. When implementing AA against the adversarially trained model on ImageNet, untargeted versions of APGD-DLR and FAB attacks are used instead, due to high computational cost.

\begin{table*}[t]
  \centering
  \caption{ASR (\%) of each attack on CIFAR-10 and ImageNet, where NAT (ADV) means the primal model is naturally (adversarially) trained.}
  \resizebox{\columnwidth}{!}{
    \begin{tabular}{cccccccccccc}
    \toprule
    \multirow{2}[2]{*}{Dataset} & Primal  & PGD   & PGD   & AA    & AA    & C\&W    & C\&W    & \multirow{2}[2]{*}{DeepFool} & \multirow{2}[2]{*}{Square} & \multirow{2}[2]{*}{TA} & \multirow{2}[2]{*}{VNI-FGSM} \\
          & Model & $\epsilon = \frac{1}{255}$ & $\epsilon = \frac{8}{255}$ & $\epsilon = \frac{1}{255}$ & $\epsilon = \frac{8}{255}$ & $\kappa=0$ & $\kappa=1$ &       &       &       &  \\
    \midrule
    \multirow{2}[4]{*}{CIFAR-10} & NAT   & 38.21  & 100.00  & 46.31  & 100.00  & 100.00  & 100.00  & 100.00  & 99.87  & 99.80  & 89.35  \\
\cmidrule{2-12}          & ADV   & 4.92  & 45.37  & 6.36  & 50.51  & 62.37  & 77.92  & 100.00  & 44.68  & 99.73  & 3.48  \\
    \midrule
    \multirow{2}[4]{*}{ImageNet} & NAT   & 95.20  & 100.00  & 100.00  & 100.00  & 100.00  & 100.00  & 100.00  & 100.00  & 99.70  & 97.70  \\
\cmidrule{2-12}          & ADV   & 10.50  & 75.50  & 13.30  & 81.50  & 30.90  & 70.70  & 100.00  & 62.00  & 99.40  & 9.24  \\
    \bottomrule
    \end{tabular}%
    }
  \label{tab:supp_attack}%
\end{table*}%

Except for TA, which is implemented using the code\footnote{\url{https://github.com/xiaosen-wang/TA}} provided by authors, all other adversarial attacks are conducted using torchattacks~\cite{kim2020torchattacks}, and the attack success rate (ASR) of each attack is shown in Table \ref{tab:supp_attack}.
It can be observed from Table \ref{tab:supp_attack} that adversarially trained models are still vulnerable to black-box attacks, and this phenomenon has also been observed and demonstrated in~\cite{dong2020benchmarking}. In addition, the performance of some detection methods can be distorted by black-box attacks~\cite{aldahdooh2022adversarial}. 
As a result, it is crucial to evaluate both detection and defense methods using the black-box attack scenario. Moreover, a detection-based module that complements defended models needs further exploration.

\subsection{Details on Detection Baselines}
\label{sec:supp_2}
Four detection methods are employed in our evaluation for comparison, and their implementation details are introduced as follows. We adjust the parameters of these detection baselines to obtain the best detection performance. Experiments are conducted using PyTorch on two NVIDIA A800 80GB PCIe GPUs.

\noindent \textbf{Feature Squeeze (FS) (Black-Box Detection).} Bit depth squeezer, local smoothing squeezer, and non-local smoothing squeezer are adopted in our work, consistent with those used in~\cite{xu2017feature}, and the implementation follows the released code\footnote{\url{https://github.com/mzweilin/EvadeML-Zoo}}. 
We reduce the original 8-bit images in test sets of two datasets to 5-bit images. Local smoothing squeezer is the median smoothing method with the $2\times2$ sliding window, where the center pixel is located at the lower right, and \emph{reflect padding} is used for pixels on the edge. These two squeezers remain invariant when implementing FS.
For the non-local smoothing squeezer, a variant of the Gaussian kernel is used. In different settings, the following parameters are applied:

\begin{itemize}
    \item For the naturally trained VGG19 on CIFAR-10, we set the search window size \(a = 13\), patch size \(b = 3\), and filter strength \(c = 2\).
    \item For the adversarially trained VGG19 on CIFAR-10, the parameters are \(a = 13\), \(b = 3\), and \(c = 4\).
    \item For the naturally trained ResNet50 on ImageNet, the parameters are \(a = 11\), \(b = 3\), and \(c = 4\).
    \item For the adversarially trained ResNet50 on ImageNet, we use \(a = 11\), \(b = 2\), and \(c = 3\).
\end{itemize}
The calculation of the probability of a test sample being an AE strictly follows~\cite{xu2017feature}.

\begin{table*}[t]
  \centering
  \caption{Comparison of AUC scores (\%) of PID by employing different metrics for quantifying the prediction inconsistency on CIFAR-10, where NAT (ADV) means the primal model is naturally (adversarially) trained.}
  \resizebox{\columnwidth}{!}{
    \begin{tabular}{ccccccccccccc}
    \toprule
    Primal & \multirow{2}[2]{*}{Metric} & PGD   & PGD   & AA    & AA    & C\&W    & C\&W    & \multirow{2}[2]{*}{DeepFool} & \multirow{2}[2]{*}{Square} & \multirow{2}[2]{*}{TA} & \multicolumn{1}{c}{\multirow{2}[2]{*}{VNI-FGSM}} & \multicolumn{1}{c}{\multirow{2}[2]{*}{Average}} \\
    Model &       & $\epsilon = \frac{1}{255}$ & $\epsilon = \frac{8}{255}$ & $\epsilon = \frac{1}{255}$ & $\epsilon = \frac{8}{255}$ & $\kappa=0$ & $\kappa=1$ &       &       &       &       &  \\
    \midrule
    \multirow{4}[8]{*}{NAT} & $I_{pred} =1-g_y(x)$ & \textbf{99.81 } & 98.54  & \textbf{99.85 } & 98.71  & \textbf{99.93 } & \textbf{99.45 } & \textbf{99.86 } & \textbf{99.88 } & \textbf{99.85 } & 97.02  & \textbf{99.29 } \\
\cmidrule{2-13}          & $I_{pred} =f_y(x)-g_y(x)$ & 99.59  & \textbf{98.80 } & 99.68  & \textbf{98.94 } & 99.52  & 97.81  & 99.48  & 97.98  & 96.70  & \textbf{97.34 } & 98.58  \\
\cmidrule{2-13}          & $I_{pred} = \Vert f(x)_{\text{top}-n} - g(x)_{\text{top}-n} \Vert_1$ & 99.44  & 97.58  & 99.53  & 97.82  & 99.67  & 99.08  & 99.62  & 99.46  & 99.40  & 95.94  & 98.75  \\
\cmidrule{2-13}          & $I_{pred} = \Vert f(x) - g(x) \Vert_1$ & 99.58  & 98.20  & 99.65  & 98.42  & 99.71  & 99.21  & 99.75  & 99.37  & 99.30  & 96.38  & 98.96  \\
    \midrule
    \multirow{4}[8]{*}{ADV} & $I_{pred} =1-g_y(x)$ & \textbf{99.87 } & \textbf{99.68 } & \textbf{99.88 } & \textbf{99.69 } & \textbf{99.91 } & \textbf{99.90 } & \textbf{99.55 } & \textbf{99.82 } & \textbf{99.82 } & \textbf{94.90 } & \textbf{99.30 } \\
\cmidrule{2-13}          & $I_{pred} =f_y(x)-g_y(x)$ & 99.34  & 97.74  & 99.37  & 97.60  & 98.30  & 98.24  & 92.23  & 98.12  & 95.94  & 90.48  & 96.74  \\
\cmidrule{2-13}          & $I_{pred} = \Vert f(x)_{\text{top}-n} - g(x)_{\text{top}-n} \Vert_1$ & 71.50  & 73.35  & 69.83  & 74.37  & 89.58  & 92.12  & 87.36  & 91.04  & 82.73  & 45.71  & 77.76  \\
\cmidrule{2-13}          & $I_{pred} = \Vert f(x) - g(x) \Vert_1$ & 66.08  & 91.16  & 63.30  & 90.58  & 84.73  & 88.49  & 86.77  & 89.60  & 76.63  & 52.06  & 78.94  \\
    \bottomrule
    \end{tabular}%
    }
  \label{tab:supp_scheme}%
\end{table*}%

\noindent \textbf{DiffPure (Black-Box Detection).} DiffPure is originally designed as a defense method, and we modify it as a black-box detection method in this work. We use the strategies described in~\cite{nie2022diffusion} to purify the inputs, following the released code\footnote{\url{https://github.com/NVlabs/DiffPure}}, where the pre-trained diffusion models Score-SDE~\cite{song2021score} and Guided Diffusion~\cite{dhariwal2021diffusion} on CIFAR-10 and ImageNet are adopted, respectively. On CIFAR-10, timesteps $t^*=0.10$ and $t^*=0.15$ are used for the naturally and adversarially trained VGG19 models, respectively. On ImageNet, $t^*=0.15$ and $t^*=0.30$ are employed for the naturally and adversarially trained ResNet50 models, respectively.

After obtaining the purified test sample $x'$ for the test sample $x$, we adopt the $\ell_1$ norm of difference of whole prediction vectors from the primal model on these two samples to calculate the probability of the test sample being an AE (denoted as $prob$), which can be described as,
\begin{equation}
    prob = \Vert f(x)-f(x')\Vert_1,
\end{equation}
where the primal model outputs confidence scores $f(x)=\{f_1(x), f_2(x), \cdots, f_k(x)\}$, $f_i(x) \in [0,1]$, and $k$ is the number of classes.

\noindent \textbf{Sensitivity Inconsistency Detector (SID) (White-Box Detection).}
SID~\cite{tian2021detecting} trains a dual classifier with the transformed decision boundary by adding the Weighted Average Wavelet Transform (WAWT) layer. The prediction differences between the primal classifier and dual classifier are used to train a detector composed of two fully connected layers. The implementation of SID can be computationally expensive on ImageNet, so we adopt it only on CIFAR-10.

We train a dual classifier for naturally and adversarially trained VGG19 models following~\cite{tian2021detecting} and the released code\footnote{\url{https://github.com/JinyuTian/SID}}, where the dual classifier consists of the same VGG19 architecture and a WAWT layer. 
For the white-box SID setting (denoted as SID*), we evaluate detection performance when the detector is trained and tested on AEs generated by the same attack against the same primal model.  
Specifically, for each attack, the corresponding AEs are split into 60\% for training, 10\% for validation, and 30\% for testing. The detector is trained for 100 epochs with a batch size of 80 and a learning rate of 0.001. The AUC score is then computed on the test set for each attack.

For the black-box SID, we use the naturally and adversarially trained VGG16 models as the primal models to mitigate performance degradation caused by the architectural differences from VGG19, ensuring a competitive baseline. The naturally trained VGG16 model achieves 92.19\% test accuracy, and the adversarially trained one, trained using PGD-8/255, achieves 79.29\% test accuracy. The dual classifier composed of the same VGG16 architecture and a WAWT layer is also trained.
AEs are then generated by attacks against these two VGG16 models, with each attack configured using the same parameters as described in Section \ref{sec:supp_1}.
The AE split and training parameters are the same as previously described to train the detector. When detecting AEs generated against the naturally and adversarially trained VGG19 models, we use a detector trained on AEs from the same attack, but generated against the naturally and adversarially trained VGG16 models, respectively. 

\noindent \textbf{Expected Perturbation Score-based Adversarial Detection (EPS-AD) (White-Box Detection).} We adopt the EPS-AD as the detection baseline on ImageNet. The implementation follows the released code\footnote{\url{https://github.com/ZSHsh98/EPS-AD}}, where the pre-trained Guided Diffusion~\cite{dhariwal2021diffusion} is adopted to estimate the expected perturbation score (EPS) of test samples and timestep $t^*=0.05$. Then, the EPS-based maximum mean discrepancy (MMD) is used as the metric to measure the discrepancy between NEs and AEs and train the detector, where the detector structure is the same as the one described in \cite{zhang2023detecting}.

For the white-box EPS-AD (denoted as EPS-AD*), we generate 10,000 $\ell_\infty$-FGSM and $\ell_2$-FGSM AEs with a perturbation size of 1/255, along with 10,000 NEs to calculate their EPSs and train the detector. When detecting AEs generated against the naturally and adversarially trained ResNet50 models, we use detectors trained using FGSM AEs generated against the same models, respectively. Detectors are trained for 200 epochs with a batch size of 200 and a learning rate of 0.002.

For the black-box EPS-AD, when detecting AEs generated against the naturally and adversarially trained ResNet50 models, we use detectors trained on FGSM AEs generated against the naturally and adversarially trained ResNet101 models, respectively. 
For the two ResNet101 models, we use pre-trained weights under natural\textsuperscript{\ref{fn:a}} and adversarial\textsuperscript{\ref{fn:b}} training settings, respectively.
The number of AEs and NEs, along with the training settings, are consistent with those described above.

\begin{table*}[t]
  \centering
  \caption{Comparison of AUC scores (\%) of PID by employing different models as the auxiliary model on CIFAR-10, where NAT (ADV) means the primal model is naturally (adversarially) trained.}
  \resizebox{\columnwidth}{!}{
    \begin{tabular}{ccccccccccccc}
    \toprule
    Primal & Auxiliary & PGD   & PGD   & AA    & AA    & C\&W    & C\&W    & \multirow{2}[2]{*}{DeepFool} & \multirow{2}[2]{*}{Square} & \multirow{2}[2]{*}{TA} & \multicolumn{1}{c}{\multirow{2}[2]{*}{VNI-FGSM}} & \multicolumn{1}{c}{\multirow{2}[2]{*}{Average}} \\
    Model & Model &  $\epsilon = \frac{1}{255}$ & $\epsilon = \frac{8}{255}$ & $\epsilon = \frac{1}{255}$ & $\epsilon = \frac{8}{255}$ & $\kappa=0$ & $\kappa=1$ &       &       &       &       &  \\
    \midrule
    \multirow{3}[6]{*}{NAT} & ResNet34\_ADV & 88.18  & 96.56  & 90.25  & 96.70  & 96.93  & 96.68  & 96.96  & 96.74  & 96.46  & 95.93  & 95.14  \\
\cmidrule{2-13}          & ViT-L/16 & \textbf{99.81 } & \textbf{98.54 } & \textbf{99.85 } & \textbf{98.71 } & \textbf{99.93 } & \textbf{99.45 } & \textbf{99.86 } & \textbf{99.88 } & \textbf{99.85 } & \textbf{97.02 } & \textbf{99.29 } \\
\cmidrule{2-13}          & CLIP-ViT-L/14 & 98.53  & 97.55  & 98.73  & 97.95  & 99.25  & 98.40  & 99.08  & 99.22  & 99.31  & 95.57  & 98.36  \\
    \midrule
    \multirow{3}[6]{*}{ADV} & ResNet34\_ADV & 86.98  & 91.46  & 87.18  & 92.12  & 97.52  & 96.01  & 97.18  & 95.06  & 97.51  & 86.47  & 92.75  \\
\cmidrule{2-13}          & ViT-L/16 & \textbf{99.87 } & \textbf{99.68 } & \textbf{99.88 } & \textbf{99.69 } & \textbf{99.91 } & \textbf{99.90 } & \textbf{99.55 } & \textbf{99.82 } & \textbf{99.82 } & \textbf{94.90 } & \textbf{99.30 } \\
\cmidrule{2-13}          & CLIP-ViT-L/14 & 98.86  & 98.75  & 99.12  & 98.69  & 99.28  & 99.18  & 98.13  & 98.97  & 99.33  & 93.43  & 98.37  \\
    \bottomrule
    \end{tabular}%
    }
  \label{tab:supp_auxiliary_cifar}%
\end{table*}%

\begin{table*}[t]
  \centering
  \caption{Comparison of AUC scores (\%) of PID by employing different models as the auxiliary model on ImageNet, where NAT (ADV) means the primal model is naturally (adversarially) trained.}
  \resizebox{\columnwidth}{!}{
    \begin{tabular}{ccccccccccccc}
    \toprule
    Primal & Auxiliary & PGD   & PGD   & AA    & AA    & CW    & CW    & \multirow{2}[2]{*}{DeepFool} & \multirow{2}[2]{*}{Square} & \multirow{2}[2]{*}{TA} & \multicolumn{1}{c}{\multirow{2}[2]{*}{VNI-FGSM}} & \multicolumn{1}{c}{\multirow{2}[2]{*}{Average}} \\
    Model & Model & $\epsilon = \frac{1}{255}$ & $\epsilon = \frac{8}{255}$ & $\epsilon = \frac{1}{255}$ & $\epsilon = \frac{8}{255}$ & $\kappa=0$ & $\kappa=1$ &       &       &       &       &  \\
    \midrule
    \multirow{3}[5]{*}{NAT} & ConvNeXt-S\_ADV & 96.33  & 97.53  & 97.06  & 98.00  & 95.82  & 96.32  & 96.24  & 96.48  & 96.81  & \textbf{98.70 } & 96.93  \\
\cmidrule{2-13}          & ViT-L/16 & \textbf{98.45 } & \textbf{98.54 } & \textbf{98.74 } & \textbf{98.90 } & \textbf{98.17 } & \textbf{98.20 } & \textbf{98.36 } & \textbf{97.80 } & \textbf{97.80 } & 98.11  & \textbf{98.31 } \\
\cmidrule{2-13}          & CLIP-ViT-L/14 & 95.15  & 95.91  & 95.83  & 96.95  & 93.94  & 94.47  & 95.37  & 94.61  & 95.86  & 95.29  & 95.34  \\
\midrule
    \multirow{3}[5]{*}{ADV} & ConvNeXt-S\_ADV & 87.17  & \textbf{97.26 } & 88.82  & 97.57  & 94.77  & 96.94  & \textbf{98.54 } & 96.45  & \textbf{99.03 } & \textbf{94.03 } & 95.06  \\
\cmidrule{2-13}          & ViT-L/16 & \textbf{95.98 } & 96.17  & \textbf{98.20 } & \textbf{97.83 } & \textbf{97.23 } & \textbf{97.11 } & 98.06  & \textbf{96.90 } & 98.34  & 92.23  & \textbf{96.81 } \\
\cmidrule{2-13}          & CLIP-ViT-L/14 & 90.98  & 96.13  & 91.52  & 96.07  & 93.55  & 94.33  & 95.98  & 93.40  & 97.42  & 93.39  & 94.28  \\
    \bottomrule
    \end{tabular}%
    }
  \label{tab:supp_auxiliary_imagenet}%
\end{table*}%

\subsection{Full Experimental Results on Metrics for Quantifying Prediction Inconsistency}
\label{sec:supp_3}
The detection performance of PID by employing different metrics for quantifying the prediction inconsistency against each attack on CIFAR-10 is summarized in Table \ref{tab:supp_scheme}.
It can be observed that using $I_{pred} = 1-g_y(x)$ helps the PID to achieve the best detection performance in most cases.
After the primal model is adversarially trained, the effectiveness of metrics $I_{pred} = \Vert f(x)_{\text{top}-n} - g(x)_{\text{top}-n} \Vert_1$ and $I_{pred} = \Vert f(x) - g(x) \Vert_1$ vary according to strengths and types of attacks, generally exhibiting a significant decline.

\subsection{Full Experimental Results on Choice of Auxiliary Models}
\label{sec:supp_4}
The detection performance of PID by employing different auxiliary models against each attack on CIFAR-10 and ImageNet is summarized in Table \ref{tab:supp_auxiliary_cifar} and Table \ref{tab:supp_auxiliary_imagenet}, respectively.
It can be observed that ViT-L/16 consistently achieves the highest AUC scores on CIFAR-10, regardless of whether it is combined with naturally or adversarially trained models. While on ImageNet, especially after the primal model is adversarially trained, ConvNeXt-S\_ADV can sometimes achieve better detection performance, which is related to its superior performance and robustness compared with traditional CNNs. 
In addition, the CLIP model exhibits lower accuracy on NEs than the other two models, leading to a higher FPR on NEs and degrading detection performance. 
Nevertheless, pre-trained foundation models like CLIP are readily accessible and require no task-specific training, making them appealing choices for deploying PID under more constrained detection scenarios, such as scenarios where the training set is completely unknown.

\end{document}